\documentclass[copyright,creativecommons]{eptcs}
\providecommand{\event}{IWS 2010} 
\usepackage{breakurl}             

\usepackage{graphicx}
\usepackage{verbatim}

\usepackage{qtree}

\usepackage{wrapfig}

\usepackage{amsmath}
\usepackage{mathrsfs}

\usepackage{txfonts}
\usepackage{pxfonts}

\usepackage{wrapfig}
\usepackage{amssymb}
\usepackage{mathrsfs}
\usepackage{stmaryrd}
\usepackage{graphicx}
\usepackage{qtree}
\usepackage[T1]{fontenc} 
\usepackage[utf8]{inputenc}
\usepackage{subfigure}
\usepackage{url}
\usepackage{tabularx}
\usepackage{longtable}

\urldef{\gravite}\url{http://gravite.labri.fr/?Projects_%2F_Grants:Porgy}

\sloppy

\usepackage{xspace}

\newcommand{\IR}{\mathcal{R}}

\newcommand\putnline[3]
{\count0 = #2 \advance \count0 by 10%
\put(#1,#2){\line(0,1){#3}}%
\put(#1,\count0){\makebox(0,0){$/$}}}

\newcommand\putaxiom[3]
{\count2 = #3 \count0 = #3 \divide \count0 by 2
\count1 = #1 \advance \count1 by \count0
\put(\count1,#2){\line(-1,0){#3}}%
\put(\count1,#2){\line(0,-1){10}}%
\advance \count1 by -\count2%
\put(\count1,#2){\line(0,-1){10}}%
}

\newcommand\putcut[3]
{\count2 = #3 \count0 = #3 \divide \count0 by 2
\count1 = #1 \advance \count1 by \count0
\put(\count1,#2){\line(-1,0){#3}}%
\put(\count1,#2){\line(0,1){10}}%
\advance \count1 by -\count2%
\put(\count1,#2){\line(0,1){10}}%
}

\newcommand\putdtriangle[3]{\count0=#2 \advance \count0 by 9%
\count1=#1 \advance \count1 by -10%
\count2=#2 \advance \count2 by 15%
\put(#1,#2){\line(-2,3){10}}%
\put(#1,#2){\line(2,3){10}}%
\put(\count1,\count2){\line(1,0){20}}%
\advance \count1 by 4%
\put(\count1,\count2){\line(0,1){10}}%
\advance \count1 by 12%
\put(\count1,\count2){\line(0,1){10}}%
\put(#1,#2){\vector(0,-1){10}}%
\put(#1,\count0){\makebox(0,0){#3}}}

\newcommand{\sema}[1] {\ensuremath{\left\llbracket #1\right\rrbracket}\xspace} 

\newcommand{\ident}{\mathit{Id}}
\newcommand{\Fail}{\mathit{Fail}}
\newcommand{\repeats}{\mathit{repeat}}

\newcommand{\ra}{\rightarrow}
\newcommand{\Ra}{\Rightarrow}

\newcommand{\R}{{\cal R}}

\ifx\thm\undefined
\newtheorem{thm}{Theorem}
\fi

\ifx\exmp\undefined
\newtheorem{exmp}[thm]{Example}
\fi

 \newtheorem{example}[thm]{Example}

\ifx\definition\undefined
\newtheorem{definition}[thm]{Definition}
\fi

\title{Strategic programming on graph rewriting systems\thanks{Research partially supported by the PORGY project (INRIA Bordeaux-Sud-Ouest and King's College London) \gravite}}

\author{Maribel Fern\'andez and Olivier Namet\\
King's College London,  Department of Informatics \\
 Strand, London WC2R 2LS, U.K.\\
\url{Maribel.Fernandez,Olivier.Namet@kcl.ac.uk}\\
}

\def\titlerunning{Strategic programming on graph rewriting systems}
\def\authorrunning{Fern\'andez and Namet}

\begin{document}
\maketitle

\begin{abstract}
  We describe a strategy language to control the application of graph
  rewriting rules, and show how this language can be used to write
  high-level declarative programs in several application areas.  This
  language is part of a graph-based programming tool  built within 
the port-graph transformation and visualisation
  environment PORGY. 
\end{abstract}

\section{Introduction}

Rewriting~\cite{Nipkow:terraa} is a computation model used in
computer science, algebra, logic and linguistics, amongst others. Its
purpose is to transform syntactic objects (words, terms, programs,
proofs, graphs, \emph{etc.}, which we will call generally
expressions), by applying rewrite rules until a suitable simplied form
is obtained.  

Given an expression and a set of rewrite rules, it is often the case that
several different rules can be applied, and the same rule can be
applied to different sub-expressions. To control the rewriting process,
a \emph{strategy} of application of rules is used.

In this paper we focus on graph
rewriting~\cite{Courcelle90,1997handbook1}.  Graphs are widely used
for describing complex structures in a visual and intuitive way, e.g.,
UML diagrams, representation of proofs, microprocessor design, XML
documents, communication networks, data and control flow, neural
networks, biological systems, etc. Graph rewriting has many
applications in specification, programming, and
simulation tools, amongst others~\cite{1997handbook2,1997handbook3}.
Several graph-transformation languages and tools have been developed,
such as PROGRES~\cite{Schurr97b}, AGG~\cite{ErmelRT97},
Fujaba~\cite{NickelNZ00}, GROOVE~\cite{Rensink03},
GrGen~\cite{GeissBGHS06} and GP~\cite{Plump09}, only to mention a few.

When the graphs are large or growing along transformations, or when
the number of graph rewriting rules is large, visualisation
becomes crucial to understand the graph
evolution. PORGY~\cite{AndreiO:icgt} is a visual environment that
allows users to define graphs and graph rewriting rules, and to
experiment with a graph rewriting system in a visual and interactive
way. For instance, one may want to apply a rule $r$ on a graph $G$ at
the positions described by the subgraph $P$ to study different
rewriting derivations for $G$.
To control the application of graph rewriting rules, PORGY uses a
strategy language. In this paper we describe this strategy language
and show how it can be used to write high-level declarative programs
in several application areas.

Strategies for rewriting have been well studied in the case of
\emph{terms}: there are languages that allow the user to specify and
apply strategies controlling the use of term rewrite rules (see for
instance~\cite{elan1, Vis01.rta,TOM-RTA07}).  For graph rewriting,
some of the languages and tools mentioned above also allow users to
guide the rewriting engine, for instance by specifying the order in
which rules will be applied; PROGRESS, Fujaba and GP offer control
structures to guide the application of rules. PORGY's strategy
language draws inspiration from these languages, and also from the
strategy languages available for term rewriting.  A distinctive
feature of our strategy language is that it allows users to describe
not only the rules that need to be applied, but also the location in the graph
where they apply, and the propagation mechanism controlling successive
applications of rules (the latter is not trivial in the case of
graphs, since there is no notion of a root, so standard term rewriting
strategies based on top-down or bottom-up traversals do not make sense
in this setting).

The syntax of PORGY's strategy language includes constructs to update
the positions where rewriting rules can apply, and to specify the way
rules should be applied.  After giving the syntax of the language, and a short, 
informal description of its semantics, in this paper we show how the language
can be used to write numerical programs, geometric applications where
visualisation plays an important role (we give a program to draw a Von
Koch fractal), and also applications to game design.

Summarising, in this paper we describe a language to define
graph rewriting systems and strategies to control the application of rules. 
Programs in this language operate on graphs, which are transformed by application
of rules at positions described explicitly in the program. We give examples to
illustrate the features of the language, and in particular we show how the
 ability to control not only the rules
and sequences of rules to be applied but also the
location in the graph where rules are applied,  allows programmers to
write concise, visual programs, in several application areas. 

In a companion paper~\cite{FN10} we give the
formal semantics of the language. For more details about the PORGY 
tool for graph visualisation and transformation, for which the strategy language 
was developed, we refer the reader to~\cite{AndreiO:icgt}.

The paper is organised as follows: In Section~\ref{sec:bg} we give a
short overview of the graph rewriting formalisms used in this paper.
We describe the strategy language in Section~\ref{Strategy} and show applications
in Section~\ref{sec:app}. In Section~\ref{sec:related} we discuss related work.
Section~\ref{sec:conc} concludes and gives some directions for future work.

\section{Background}
\label{sec:bg}


A graph rewriting system is a set of rewrite rules
of the form $L \Ra R$ where $L$ and $R$ are graphs.  
Such a rule
applies to a graph $G$ if $G$ contains at least one instance of the
left-hand side $L$, i.e. a subgraph $A$ isomorphic to $L$. Each rule
specifies an interface that is used in order to rewrite $A$ in $G$:
$G$ rewrites to a new graph $G'$ obtained by replacing the instance
$A$ of $L$ by an instance $B$ of $R$, where edges that were connected
to nodes in $A$ are connected to $B$ as specified by the rule's
interface.  This rewriting process induces a transitive relation on
graphs. Each rule application is a rewriting step and a derivation is
a sequence of rewriting steps, that we will sometimes call a
computation, referring to application of rewriting to programming
languages.

There are several formal definitions of graph rewriting systems,
depending on the kind of graphs and rewriting rules that can be
defined in the system (see, for instance,
\cite{BarendregtHP:tergr,Plump98termgraph,Barthelmann96howto,LafontY:intn}).
As a particular example of graph rewriting formalism of interest, in
this paper we consider port graph rewriting, of which interaction
nets~\cite{LafontY:intn} are a particular case.  We recall below the
main notions of port graph rewriting, and refer to~\cite{AndreiK07a}
for more details.

\paragraph{Port graphs.}
A port graph~\cite{AndreiK08c} is a graph where nodes have explicit
connection points called {\em ports} and the edges are attached to
specific ports of nodes. Each edge connects at most two ports.
 Nodes and ports are labelled using 
 $\mathscr{N}$ and $\mathscr{P}$, which are two disjoint sets of node names and
port names respectively. A {\em p-signature} over $\mathscr{N}$ and
$\mathscr{P}$ is a mapping $\nabla : \mathscr{N} \ra 2^{\mathscr{P}}$
which associates a set of port names to a node name. 
Without loss of generality, we assume that the 
port names associated to different node names are different. In addition, 
ports may have state information, which is formalised using a mapping from 
port names to port states.

 Let $G$ and $H$
be two port graphs over the same p-signature.  A {\em port graph
  morphism} $f:G\ra H$ maps elements of $G$ to elements of $H$
preserving sources and targets of edges, as well as node names and
associated port name sets.

A port graph rewrite rule $L \Ra R$ is itself represented as a port
graph consisting of two port graphs $L$ and $R$ over the same
p-signature and one special node $\Ra$, called {\em arrow node}
connecting them.  $L$ and $R$ are called the {\em left-} and {\em
  right-hand side} respectively. The arrow node is used to specify the
correspondence between free ports in the left- and righ-hand sides of
the rules, i.e., the interface of the rule.  For each port $p$ in $L$
to which corresponds a non-empty set of ports $\{p_1,\ldots, p_n\}$ in
$R$, the arrow node has a unique port $r$ and the incident directed
edges $(p,r)$ and $(r,p_i)$, for all $i=1,\ldots,n$; all ports from
$L$ that are deleted in $R$ are connected to the {\em black hole} port
of the arrow node. When the correspondence between the ports in the
left- and right-hand sides is obvious we omit the ports and edges
involving the arrow node.

We illustrate the notions of port graph and port graph rewriting rule 
with examples in Section \ref{sec:app}.

Let $L\Ra R$ be a port graph rewrite rule and $G$ a port graph such
that there is an injective port graph morphism $g$ from $L$ to $G$;
hence $g(L)$ is a subgraph of $G$.  Replacing $g(L)$ by $g(R)$ and
connecting it with the rest of the graph as specified by the interface
of the rule, we obtain a port graph $G'$ which represents a result of
{\em one-step rewriting} of $G$ using the rule $L\Ra R$, written
$G\ra_{L\Ra R}G'$. There may be different injective morphisms $g$ from
$L$ to $G$; they are built as solutions of a {\em matching} problem
from $L$ to a subgraph of $G$.  If there is none, we say that $G$ is
{\em irreducible} by $L\Ra R$.  Given a finite set $\R$ of rules, a
port graph $G$ {\em rewrites} to $G'$, denoted by $G\ra_{\R}G'$, if
there is a rule $r$ in $\R$ such that $G\ra_{r}G'$.  This induces a
transitive relation on port graphs. Each {\em rule application} is a
rewriting step and a {\em derivation}, or computation, is a sequence
of rewriting steps.  A port graph on which no rule is applicable is in
\emph{normal form}.  Rewriting is intrinsically non-deterministic
since it may be possible to rewrite several subgraphs of a port graph
at the same time with different rules or the same one at different
places, possibly getting different results.

\paragraph{Interaction nets.}

Interaction nets were introduced by Lafont in 1990~\cite{LafontY:intn}
and later used as a target language for implementation of efficient
$\lambda$-calculus evaluators (see for instance~\cite{GonthierG:geoolr, MackieIC:efflei}).

\begin{figure}[!b]
\centering
\includegraphics[width=1\columnwidth, keepaspectratio]{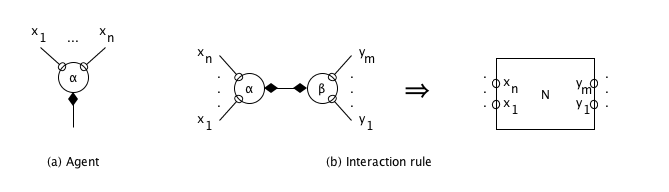}

 \caption{General format of an agent and an interaction rule}
 \label{INets}
\end{figure}

A system of interaction nets is specified by a set $\Sigma$ of symbols
with fixed arities, and a set $\IR$ of interaction rules.  An
occurrence of a symbol $\alpha\in\Sigma$ is called an \emph{agent}. If
the arity of $\alpha$ is $n$, then the agent has $n+1$ \emph{ports}: a
\emph{principal port} depicted by an arrow, and $n$ \emph{auxiliary
  ports}. Figure~\ref{INets}(a) shows a typical drawing of such an
agent. Intuitively, a net $N$ is a port graph (not necessarily
connected) where the nodes are agents.  The ports that are not connected
to another agent are said to be \emph{free}.
The \emph{interface} of a net is its set of
free ports.

Interaction rules are port graph rewrite rules where the left-hand
side consits of two agents $(\alpha,\beta)\in \Sigma\times\Sigma$
connected together on their principal ports (an \emph{active pair} or
\emph{redex}) and the right-hand side is a net $N$ with the same
number of free ports as the left-hand side. In addition, at most one
rule can be given for each pair of agents.  The diagram depicted in
Fig.~\ref{INets}(b) shows the format of interaction rules ($N$ can be
any net built from $\Sigma$).

Interaction rules can be seen as a particular kind of port graph
rewriting rules with constraints that ensure good computational properties:
pattern-matching is particularly easy in interaction nets, since
  patterns are graphs containing just two nodes, and the
  transformations are local and strongly confluent.

 In some contexts, for instance when using interaction rules to
 program functions, we need a weaker notion of normal form corresponding to
the notion of  weak head normal form in $\lambda$-calculus. Interface
 normal form, defined in~\cite{FernandezM:calin}, can be computed by
 a strategy which applies rules at the interface (if possible) in order to
 minimise the length of the reduction sequence. We give 
 examples below.

\section{Strategy language}
\label{Strategy}

Since the application order of a set of port graph rewrite rules on a
port graph may alter the result and the resources (e.g., time and
space) needed to reach the result, a rewriting strategy is needed to
control the rewriting process.  As mentioned in the introduction,
strategic rewriting has been well studied for \emph{term} rewriting
systems (see, e.g.,~\cite{BorovanskyKKMR98,Vis01.rta,TOM-RTA07}).
Here we present a strategy language for \emph{graph} rewriting
systems, where strategies take into account rule selection and
position selection in a graph. 

The notion of position in a graph is rather delicate since there is no
notion of root. Hence standard term rewriting strategies based on
top-down or bottom-up traversals do not make sense in the case of graphs. A
specific language is needed to define strategies for graph
rewriting systems.
We propose a strategy language where expressions include operators to
select rules and the positions where the rules are applied, and also
to change the positions along the derivation.

Let $G$ be the graph to be
rewritten. Then a \emph{position} $P$ is a subgraph of $G$ specifying 
where rewrite rules are allowed to apply. More precisely, we require
that the homomorphic image of the left-hand side of the rule $(L \Ra R)$
has a non-empty intersection with this subgraph, that is at least one node in $g(L)$ is in $P$. A simple and
intuitive example is to define $P$ to be a specific node and in this case,
only the rules having this node in the homomorphic image of
 their left-hand sides are allowed to apply.

We call the structure $G[P]$ consisting of a graph and a position 
a \emph{located graph}. A \emph{program} is composed of a located graph 
and a strategy expression. Programs act on graphs; the result of a terminating
program is a new graph.

The syntax of the strategy language is given in
Figure~\ref{fig:syntax-strategies}, where we show the grammars $S$,
$A$ and $T$  for generating \emph{strategy
  expressions, rule applications and position transformations}, respectively.  
The main idea is that given a set of
graph rewrite rules, an initial located graph and a strategy
expression in this language, rewriting sequences (i.e., derivations)
are generated, starting from the initial graph, by applying the rules as
specified by the strategy expression.  The most basic expressions
generated by the grammar are $\ident$ and $\Fail$, 
namely identity and failure.
A position transformation (see the grammar for $T$ in Figure~\ref{fig:syntax-strategies})
 applied on a located graph $G[P]$ only affects the
subgraph specifying the positions (i.e., $P$).
An application (see the grammar for $A$  in Figure~\ref{fig:syntax-strategies}) 
on a located graph may change both the graph and the positions.
A strategy expression embeds the previous constructs and combines them using
sequential composition, iteration and conditional choice 
(see the grammar for $S$  in Figure~\ref{fig:syntax-strategies}).
Below we describe the constructs of the strategy language and their effect on 
a given located graph; we finish this section with some examples.

\begin{figure}[!t]
\centering
\fbox{
\renewcommand{\arraystretch}{1.5}
\begin{tabular}{lrrrl}
 & & & & Let G: Graph, P:Position, $\rho$:Property, m,n:Int \\
  {\bf (Position } & & $T$ & $:=$ & $\mathit{CrtPos} \mid
  \mathit{AllSuc} \mid \mathit{OneSuc} \mid \mathit{NextSuc}
  \mid \mathit{SetPos}(P)$ \\ 
  {\bf \ \ Transformation)} & & & $\mid$ &  $ \mathit{Property}(\rho,G)
  \mid  T \cup T \mid 
  T \cap T \mid \overline{T} \mid T - T$ \\
  {\bf (Application)} & & $A$ & $:=$ & $\ident \mid \Fail \mid (L_i \Ra R_i)_P
  \mid A\parallel A \mid A\interleave A \mid A^{\parallel(m,n)} $ \\
  {\bf (Strategy)} & & $S$ & $:=$ & $  T \mid A \mid S;S  \mid
  S+S \mid \mathit{ppick}(S,\ldots ,S) $ \\
  & & & $\mid$ & $
  \mathit{while}~(S)~\mathit{do}~(S)~\mathit{min}(m)~\mathit{max}(n)$
  \\ 
  & & & $\mid$ & $ \mathit{if}~(S)~\mathit{then}(S)~\mathit{else}(S) 
  \mid \mathit{PnotEmpty} \mid  \langle S\rangle  $
\end{tabular}
}
\caption{Syntax of the strategy language.}
\label{fig:syntax-strategies}
\end{figure}

\smallskip
\noindent 
{\bf Position transformations.} These allow programmers to focus on
different parts of the graph during the rewriting process.  A position
transformation $T$ applies to a located graph $G[P]$ to produce a new
graph $G[P']$. The new position $P'$ computed by $T$ is defined as
follows.  $\mathit{CrtPos}$ returns the current position in the
located graph (it is akin to the identity transformation).
$\mathit{AllSuc}$ returns immediate successors of
all nodes of $P$, where
an immediate successor of a node A is a node that has a port  connected to a port of A.
$\mathit{OneSuc}$
looks for all the immediate successors of all nodes in $P$ and  picks
one of those randomly.  
$\mathit{NextSuc}$  computes  successors of nodes in $P$ using a designated
port for each node (for Interaction Nets, this is the principal port).
$\mathit{SetPos}(P')$ changes the position
$P$ to $P'$, where $P'$ is a subgraph of $G$ explicitly defined (for instance 
by  selecting nodes through PORGY's visual interface). 
$\mathit{Property}(\rho,G')$ updates $P$ to contain only the
nodes from the graph $G'$ that satisfy the property $\rho$  ($G'$ will generally
 be the given located graph or the subgraph $P$).
The set theory operators $union$, $intersection$, $complement$ and
$subtraction$ apply to positions since they are graphs considered as 
sets of nodes and edges.

\smallskip
\noindent 
{\bf Applications.}  The application of $\ident$ never fails and
leaves the graph unchanged whereas $\Fail$ always fails (it leaves the
graph unchanged and returns failure). $(L \Ra R)_M$ represents the
application of the rule $L \Ra R$  in $G[P]$ if $L \cap P$ is not empty (i.e., the 
graph reduced must overlap with the position $P$ specified in the located graph); $M$ is the subgraph of $R$ that is added
to position $P$ (in the copy of $R$ added to $G$ in the rewrite step, the nodes in $M$ become part of $P$; the rest of $R$ is only added to $G$
and not to $P$).  $A\parallel A'$ represents simultaneous application of
$A$ and $A'$ on disjoint subgraphs of $G$ and returns $\ident$ only if
both applications are possible and $\Fail$ otherwise. $A\interleave
A'$ is a weaker version of $A\parallel A'$ as it returns $\ident$ if
at least one application of $A$ or $A'$ is possible.
$A^{\parallel(m,n)}$ applies $A$ simultaneously a minimum of $m$ and a
maximum of $n$ times. If the minimum is not satisfied then $\Fail$ is
returned and $\ident$ otherwise. If $n$ is a negative integer then no
maximum is considered.

\smallskip
\noindent
{\bf Strategies.}  
The expression $S;S'$ represents sequential application of $S$ followed by $S'$, and
$S+S'$ implements McCarthy's AMB operator \cite{Mccarthy63abasis}.
%
The strategy $\mathit{ppick}(S,S')$ is a weaker version of $S+S'$ as it randomly
picks one of the strategies for application. 
For iterations, we have expressions of the form 
$\mathit{while}~(S)~\mathit{do}(S')~\mathit{min}(m)~\mathit{max}(n)$
which keep on sequentially applying $S'$ while the expression $S$ evaluates to $Id$;
if the minimum of $m$ successful applications of $S'$ is not satisfied
then it returns $\Fail$ or else $\ident$ is returned.  Similar to
$A^{||(m,n)}$, setting $n$ to a negative integer eliminates the
maximum.
The strategy $\mathit{if}~(S)~\mathit{then}(S')~\mathit{else}(S'')$ checks if the
application of $S$ to $G[P]$ returns $Id$ in which case $S'$ is applied to $G[P]$
otherwise $S''$ is applied ($S$ is checked on a copy of $G$ and not on $G$ itself so the graph is not affected during the checking process). 
$\mathit{PnotEmpty}$ returns $\Fail$ if $P$ is empty and $\ident$
otherwise. 
This 
can be used  for instance inside the condition of an $\mathit{if}$ or $\mathit{while}$, 
to check if the strategy makes $P$ empty or not, instead of checking if the strategy itself can be applied.
The strategy $\langle S\rangle$ applies $S$ and considers $S$ as one atomic rewriting
step in the derivation tree. This is useful to abstract  several
reduction steps as  one for visualisation purposes.

The semantics of the strategy constructs defined by the grammars in
Fig.~\ref{fig:syntax-strategies} has been formally defined
in~\cite{FN10} using rewrite rules 
that apply to  programs $\sema{S, G[P]}$ where $S$ is a
strategy and $G[P]$ a located graph. 
These rewrite rules  are governed by a top-down, left-right meta-strategy, which ensures
the following property (we refer the reader to~\cite{FN10} for details and proofs).

{\bf When the strategy $S$ is grammatically correct, 
an expression $\sema{S, G[P]}$  either rewrites
indefinitely (when the strategy $S$ does not terminate) or rewrites to
an expression of the form $\sema{E,G'[P']}$,  where $E$ is $\ident$ or $\Fail$ and $G'[P']$ is a new located graph. }

We will  now  illustrate the strategy  language with a few examples.

\begin{example}
In many applications, we need to repeatedly apply a rule (or set of
rules, or more generally, we need to iterate a strategy) as long as possible. 
This can be easily done in the language by defining the expression $\repeats_*(S)$ 
 as shown below. The expression $\repeats_+(S)$ applies $S$  at least once.
\begin{itemize}
\item $\repeats_*(S)\triangleq
  \mathit{while}(S)~\mathit{do}~(S)~\mathit{min}(-1)~\mathit{max}(-1)$
\item $\repeats_+(S) \triangleq
  \mathit{while}(S)~\mathit{do}~(S)~\mathit{min}(1)~\mathit{max}(-1)$
 \end{itemize}

More concretely, the following expression can be used if $(A
  \ra B)$ and $(C \ra D)$ are two chemical reactions that must take
  place together (represented as graph rewriting rules).
$$\repeats_{*}((A \Ra B)|| (C \Ra D))$$

Other well-known strategy operators, such as $Not$, $orelse$ and $Try$ are defined below, together with a strategy to compute interface normal forms of interaction nets.
\begin{itemize}
\item $\mathit{Not}(S) \triangleq
  \mathit{if}(S)~\mathit{then}(\Fail)~\mathit{else}(\ident)$ is a strategy that fails if $S$ succeeds, and conversely, it succeeds if $S$ fails.
\item $S~ orelse ~S' \triangleq
  \mathit{if}(S)~\mathit{then}(S)~\mathit{else}(S')$, applies
  $S$ if possible, otherwise it applies $S'$ and
  fails if neither strategy is applicable.

\item $\mathit{Try}(S) \triangleq
  \mathit{if}(S)~\mathit{then}(S)~\mathit{else}(\ident)$ is a
  strategy that behaves like $S$ if $S$ succeeds, but if $S$ fails
  then it behaves like $\ident$.
\item  The interface of a graph $G$  is the set of nodes of $G$ 
that have a free port. They can
  be selected by defining the position $Property(interface, G)$,
  denoted $Int$. Then, if $G$ is an interaction net and $P$ its interface,
the program  $\sema{\repeats_*((R_1 ; \mathit{Int}) \mathit{orelse}~
  \mathit{Next}),G[P]}$  computes the  Interface Normal Form 
   of $G$ with respect to $R_1$.

\end{itemize}

\end{example}

The strategy language described in this section has been implemented
into PORGY in the form of a plug-in.  The user types in a strategy
expression which is first parsed by an algorithm to create a
\emph{strategy tree}. A \emph{strategy engine} then takes the tree and
applies a series of rewrite rules on the tree, following the
meta-strategy defined by the formal semantics of the language in~\cite{FN10}.  Some
of these rules will call for an application or a transformation to be
performed on the graph. The strategy engine terminates when the tree
is reduced to only its root which will either be $Id$ or $Fail$ (a
successful strategy or a failed one, respectively). The user will then
have a visual representation of the final state of the graph as well
as a step by step trace of the strategy  applied.

\section{Applications} 
\label{sec:app}
\subsection{Arithmetic programs with Interaction Nets}

\begin{figure}
\centering
\includegraphics[width=1\columnwidth, keepaspectratio]{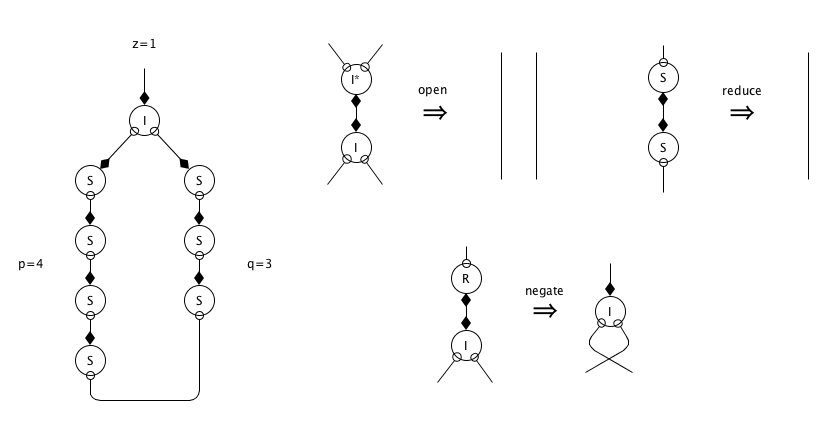}
\caption{An example number and the \emph{open}, \emph{reduce} and \emph{negate} rules.}
\label{arithmetic1}
\end{figure}

In term rewriting systems with a finite signature, natural numbers are
often represented using two function symbols: $S$ and $0$. Then the
number $n$ is represented by a term $S(S(\ldots S(0)\ldots )$ with $n$
occurrences of $S$. This representation is inefficient, but
in~\cite{MackieIC:hab} it is shown that using Interaction Nets we can
implement efficiently arithmetic operations on integers, with a finite
signature, by representing a number $z$ in the form of a difference
list $p-q$.  The $I$ agent is used as $head$ of a number, and holds two
lists of $S$ agents: a left list containing $p$ and a right list
containing $q$; see Figure \ref{arithmetic1} for an example of the
number $1$ represented as $4-3$. We also note that there are infinite 
representations for each number in this way:
$1=4-3=6-5=7-6=...$

The $open$ rule given in Figure \ref{arithmetic1} extracts both lists from a number so that they can be used for arithmetic operations. If two lists are put head to head, the $reduce$ rule will eventually return a single list containing the absolute value of the difference of the lists. The right-hand sides of these two rules are depicted as wires, that is, when the rule $reduce$ is applied, the ports connected to the $S$ agents being reduced will be connected together, and similarly for the $open$ rule. The $negate$ rule switches the left and right list of a number, giving us its negative. The $negate$ rule has its entire right hand side in its subgraph $M$, so if $P$ is initially the whole graph then no matter which of these three rules we apply, $P$ will always be the entire graph.

\begin{figure}
\centering
\includegraphics[width=1\columnwidth, keepaspectratio]{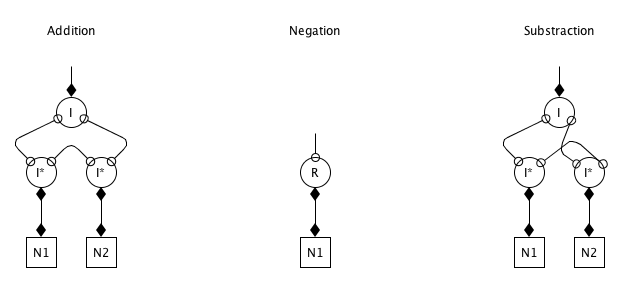}
\caption{Modelling Addition, Negation and Subtraction.}
\label{arithmetic2}
\end{figure}

Using these three rules we can model $Addition$, $Negation$ and $Subtraction$, as seen in Figure~\ref{arithmetic2}.
If we liken the size of a graph to memory space, we could then  prioritise the $reduce$ rule so that the graph is always kept at its smallest. A useful strategy to design would then be:

\centerline{ArithStrategy: repeat$_*$(repeat$_*$($reduce$);Try($negate$);Try($open$))}

Here, we apply first the rule $reduce$ as much as possible, to simplify the representation of the numbers, followed by applications of negate and open to perform the arithmetic operations.

\subsection{Von Koch Fractal}

\begin{figure}
\centering
\includegraphics[width=1\columnwidth, keepaspectratio]{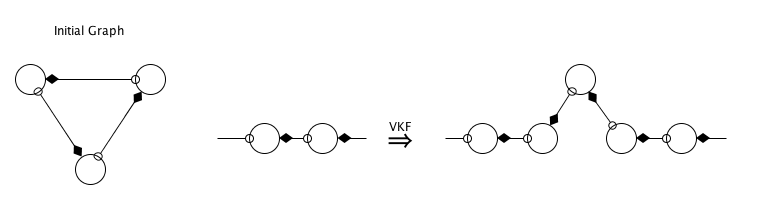}
\caption{Modelling the Von Koch Fractal.}
\label{vonkoch1}
\end{figure}

To draw a Von Koch Fractal (see Figure \ref{vonkoch1}), we only need
one agent type and one rule. Our initial graph is a triangle
and has one (and only one) of its nodes in P. We define
a rule \emph{vonKoch} of type $(L \Ra R)_M$ (see \emph{VFK} in
Figure~\ref{vonkoch1}) such that $M$ contains the right-most agent from
the right-hand side of the rule.
This means that after each application of the rule \emph{vonKoch} the subgraph $P$ 
is updated to restrict the next application of rules to the neighbouring segment.
Visually, our rule will \emph{travel} round the triangle, segment by segment,
 gradually creating a more complex fractal after each round trip.

\begin{figure}
\centering
\includegraphics[width=0.5\columnwidth, keepaspectratio]{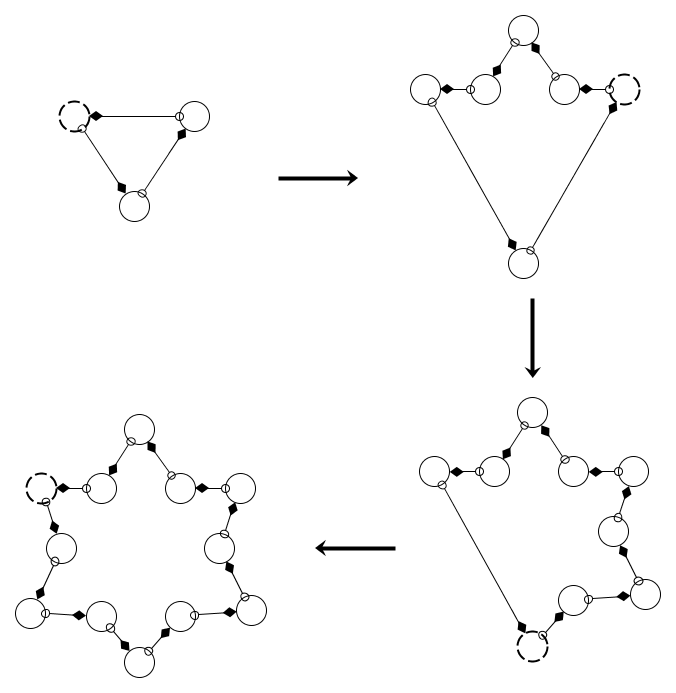}
\caption{The Von Koch Fractal.}
\label{vonkoch2}
\end{figure}

In Figure \ref{vonkoch2}, we can see three successive applications of \emph{vonKoch}. Agents drawn with dashed lines are agents that are in P.
The VKF strategy used is simply:

\centerline{ while($vonKoch$)do($vonKoch$)min(0)max($m$)}
where $m$ is the number of iterations required.

Without the notion of position in the strategy language, the application of the $VKF$ rule would have been random and the fractal generated would not necessarily be balanced. 
The strategy language allows us to define for each rewrite rule the nodes in the right-hand side that will become part of the position subgraph. The ability to update positions directly 
is exploited in this example to obtain a simple and concise program that goes round the triangle creating the fractal in a balanced way.

For this example, it is important that the shape and layout of the right hand side of the rules is preserved during the rewriting step to ensure the proper shape of the fractal. In PORGY the user can specify  for each rule whether the shape of the right hand side is preserved or not.

\subsection{Game example: Pacman}

\begin{figure}
\centering
\includegraphics[width=1\columnwidth, keepaspectratio]{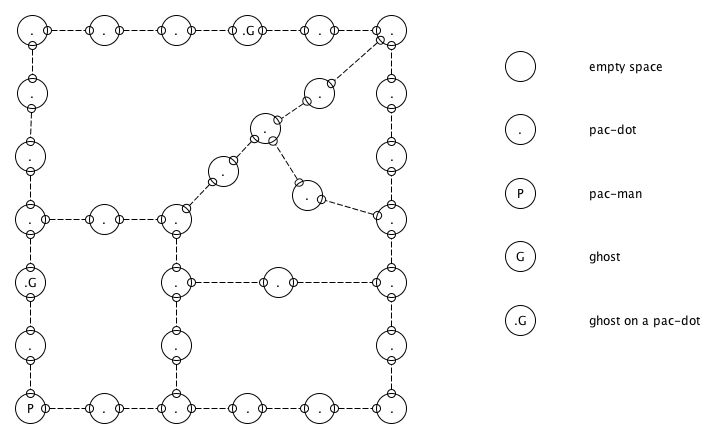}
\caption{The pac-man playing field.}
\label{initpac}
\end{figure}

To simulate a game of pac-man, we use the initial graph in Figure
\ref{initpac} with the five types of nodes depicted.  We assume all
nodes in this system to have four ports each, one for each
direction. For visual simplicity we will not draw any free ports or
ports whose state does not affect a rewrite rule.

\begin{figure}[!t]
\centering
\includegraphics[width=0.9\columnwidth, keepaspectratio]{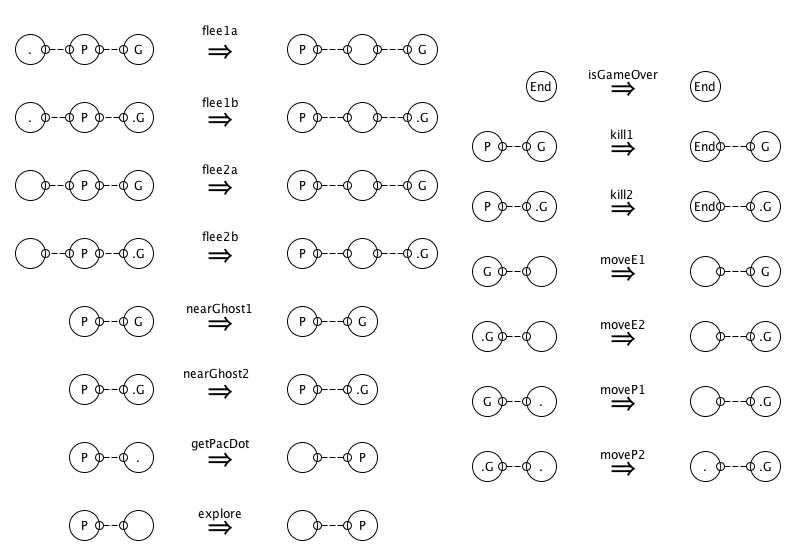}
\caption{The set of rules to control pac-man (left) and to control the ghosts (right).}
\label{pacrule}
\end{figure}

The rewrite rules for pac-man can be found in Figure \ref{pacrule}. These rules, with the help of a strategy, will help simulate a basic behavior for pac-man. Pac-man's first instinct will be to flee any nearby ghosts (rules \emph{flee1a}, \emph{flee1b}, \emph{flee2a} and \emph{flee2b}). If pac-man is not near any ghosts he will then seek out pac-dots (rule \emph{getPacDot}) and then if not near any pac-dots, he will proceed to explore the level (rule \emph{explore}).

The strategy for controlling pac-man is as follows:

\begin{itemize}

\item pacAI: if($nearGhost1$ orelse $nearGhost2$) then (Flee) else (Move)

\item Flee: if($flee1a$ orelse $flee1b$) then ($flee1a$ orelse $flee1b$) else (Try($flee2a$ orelse $flee2b$))

\item Move: if($getPacDot$) then ($getPacDot$) else (Try($explore$))

\end{itemize}

The rewrite rules for the ghosts can be found in Figure \ref{pacrule}. Like for pac-mac, these rules and a set of strategies will help simulate the behaviour of the ghosts. A ghost's first priority is to eat pac-man (rules \emph{kill1} and \emph{kill2}). If pac-man is not nearby, then a ghost will try move to a space with no pac-dots (rules \emph{moveE1} and \emph{moveE2}) since following empty spaces should lead the ghost to pac-man. If a ghost can only move to a space with a pac-dot then do so (rules \emph{moveP1} and \emph{moveP2}).

The strategy for controlling ghosts is as follows:

\begin{itemize}

\item ghostAI: if($kill1$ orelse $kill2$) then ($kill1$ orelse  $kill2$) else (gMove)

\item gMove: if($moveE1$ orelse  $moveE2$) then ($moveE1$ orelse $moveE2$) else (Try($moveP1$ orelse  $moveP2$))

\end{itemize}

The overall strategy called \emph{gameLoop} that controls the game is as follows:
We must first check that pac-man has not been eaten (by checking for the existence of a node of type \emph{End}). We then call \emph{pacAI} followed by \emph{ghostAI} for each ghost (we do this by adding pacman and all ghosts to $P$ at the start of each game loop using Property(Y,G) and making sure all the rules that involve ghosts have an empty $M$. This means every time a ghost performs an action,which removes the ghost from P, it cannot perform another one till the next game loop).

\begin{itemize}

\item gameLoop: repeat$_*$(Property(Y,G);if($isGameOver$)then($Fail$)else($pacAI$;repeat$_*$($ghostAI$))

\item Y is: type=="ghost" or "pac-man" or "End"

\end{itemize}

As we can see, a basic pac-man game does not require many agents and just six relatively 
simple strategies are sufficient to model it using our language.

Adding a scoring system would be trivial: each time pac-man eats a
pellet, a \emph{point} agent would be created and added to a
\emph{list} of points which can then be counted at the end of a game.

If there is more than one possible application of a rule, the implementation will pick one of the possibilities at random. This will create a different game each time.

\subsection{Labyrinth}

\begin{figure}[t]
\centering
\includegraphics[width=0.9\columnwidth, keepaspectratio]{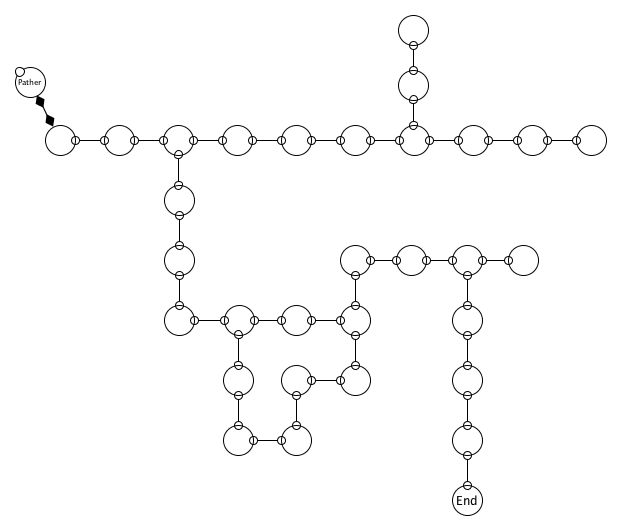}
\caption{An example of a labyrinth.}
\label{labyrinthMap}
\end{figure}

We will now give a program to find a path  in a labyrinth. The labyrinth is represented as a graph built out of \emph{Labyrinth} agents, as shown in Figure~\ref{labyrinthMap}, where \emph{Labyrinth} agents are depicted as empty circles and exits are represented with an \emph{End} agent. The initial located graph in this example has a \emph{Pather} agent connected to the start of the maze.

\begin{figure}[h]
\centering
\includegraphics{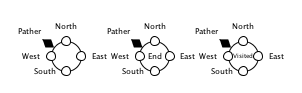}
\caption{A \emph{Labyrinth} agent, $End$ agent and $Visited$ agent.}
\label{labyrinthPosition}
\end{figure}

A \emph{Labyrinth} agent has five ports, one for each cardinal direction North, East, South and West and a \emph{Pather} port, where a \emph{Pather} agent can attach to (see Figure \ref{labyrinthPosition}). The $End$ agent has the same ports as a \emph{Labyrinth} agent but will react differently when a $Pather$ agent is connected to it. We will also have a $Visited$ agent, which has the same ports as a \emph{Labyrinth} agent but like the $End$ agent, will react differently when a $Pather$ agent connects to it. Lastly we have a $PATH$ agent which has the same ports as the \emph{Labyrinth} agent and will be used to replace \emph{Labyrinth} agents so that a visible path will be drawn from the start to the exit of the labyrinth.

For the sake of clarity, in the following diagrams the four directional ports will not be labelled but will be drawn in the standard orientation on the agents. In the rules, a white port means that the port must be connected, a crossed port must be free and a black port means either connected or free.

A \emph{Pather} agent has a \emph{Position} port and a \emph{List} port. The $Position$ port connects to a \emph{Labyrinth} agent and the $List$ port will connect to a list of $Direction$ agents (representing the path followed so far). We have four $Direction$ agents $N$ , $E$, $S$ and $W$ that each have two ports: a $Next$ and a $Prev$ port. We will also need a $Drawer$ agent (with the same ports as a $Pather$ agent) that will travel back to the start of the labyrinth, following a list of $Directions$ and replacing \emph{Labyrinth} agents with $PATH$ agents.

\begin{figure}[h]

\centering
\includegraphics{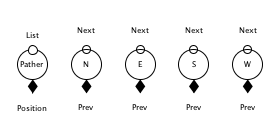}
\caption{A \emph{Pather} agent and the four \emph{Direction} agents.}
\label{labyrinthPatherDirection}
\end{figure}

\begin{figure}[h]

\centering
\includegraphics[width=0.9\columnwidth, keepaspectratio]{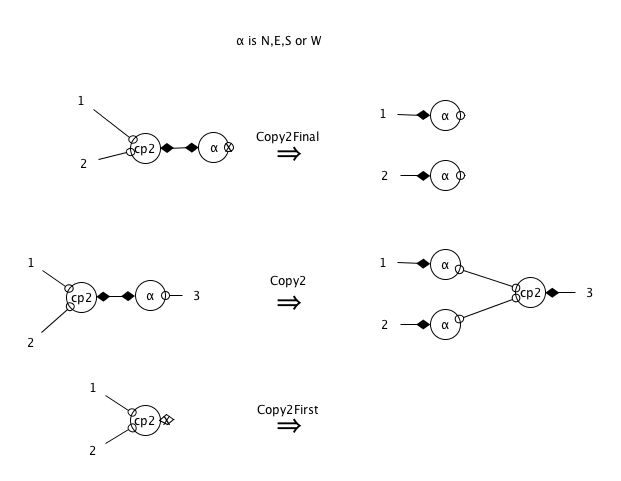}
\caption{The set of rules for $cp2$.}
\label{labyrinthCopyRules}
\end{figure}

\begin{figure}[h]

\centering
\includegraphics[width=0.5\columnwidth, keepaspectratio]{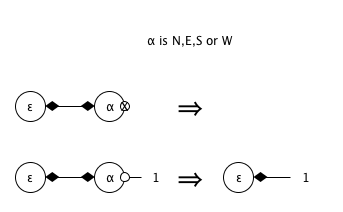}
\caption{The set of rules for $\epsilon$.}
\label{labyrinthEat}
\end{figure}

Lastly, we have some $management$ agents: two $copy$ agents ($cp2$ and $cp3$) and a $delete$ agent named $\epsilon$. The $copy$ agents take a list of directions and duplicate ($cp2$) or triplicate ($cp3$) it. The $\epsilon$ agent takes a list and deletes it. The rewrite rules for $cp2$ can be found in Figure \ref{labyrinthCopyRules} (the $cp3$ rules are similar  but produce  three copies) and the rewrite rules for $\epsilon$ in Figure \ref{labyrinthEat}.

The program consists of the  strategy expression $LabStrat$ described below, and a located graph representing the labyrinth, where the only node in the initial subgraph $P$ is the $Pather$ agent marking the starting point in the labyrinth. The strategy has two main parts, which we call Step 1 and Step 2.  Step 1  attempts to find a path to the exit of the labyrinth, by moving the starting $Pather$ until a $Pather$ agent positions itself onto the $End$ agent. If a $Pather$ agent is positioned onto an $End$ agent, a path was found and the program will move onto the Step 2. When a $Pather$ agent moves to a new position, it changes the \emph{Labyrinth} agent it moved from into a $Visited$ agent. This will ensure the $Pather$ agent never backtracks.

\begin{figure}[h]
\centering
\includegraphics[width=0.5\columnwidth, keepaspectratio]{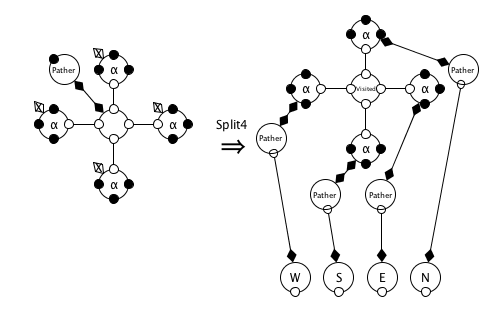}
\caption{The $split4$ rule. $\alpha$ is a \emph{Labyrinth} or $End$ agent.}
\label{labyrinthSplit4}
\end{figure}

The strategy will start by checking if a $Pather$ agent is connected to four \emph{Labyrinth} or $End$ agents that have their $Pather$ port free (rule $split4$ in Figure \ref{labyrinthSplit4}, a special case of when the starting $Pather$ is put on such a \emph{Labyrinth} agent). This rule will remove the $Pather$ agent and create four new $Pather$ agents for each of the four positions and give each one of the new $Pathers$ a corresponding $Direction$ agent (to remember the step done).

\begin{figure}[h]
\centering
\includegraphics[width=0.5\columnwidth, keepaspectratio]{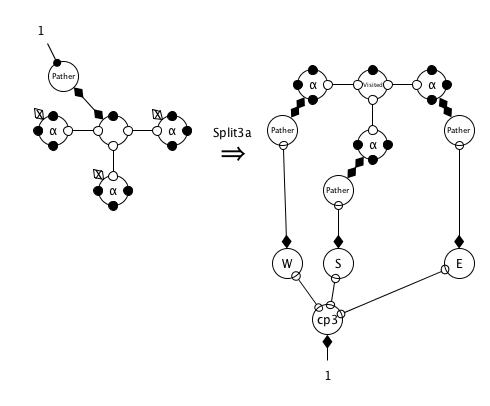}
\caption{The $split3a$ rule. $\alpha$ is a \emph{Labyrinth} or $End$ agent.}
\label{labyrinthSplit3}
\end{figure}

If $split4$ cannot be applied, the strategy will try to apply one of the four $split3$ rules $split3a$, $split3b$, $split3c$ or $split3d$. This rule deletes the original $Pather$ agent and creates three new $Pather$ agents, adding a corresponding $Direction$ agent to each of their lists, and copying the original $Pather$ agent's list onto the end of the new $Pather$ agents' lists. See Figure \ref{labyrinthSplit3} for the $split3a$ rule (the other three split3 rules are similar, taking into account the remaining combinations).

\begin{figure}[h]
\centering
\includegraphics[width=0.5\columnwidth, keepaspectratio]{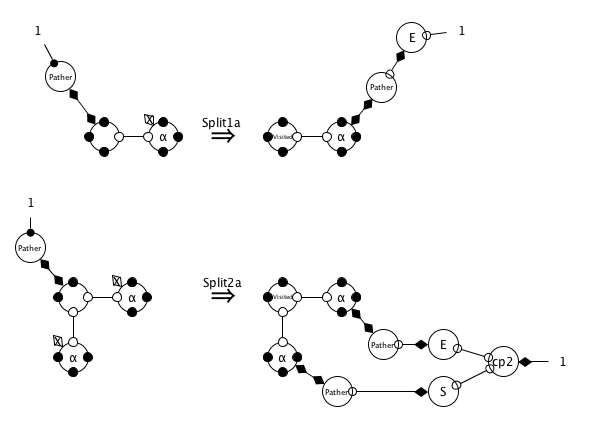}
\caption{The $split2a$ and $split1a$ rule. $\alpha$ is a \emph{Labyrinth} or $End$ agent.}
\label{labyrinthSplit}
\end{figure}

If none of the $split3$ rules can be applied, the strategy then tries all six of the $split2$ rules and if none of the $split2$ rules can be applied, it tries one of the four $split1$ rules. These rules do the same thing as the $split3$ rules but only split to two and one \emph{Labyrinth} agents respectively. See Figure \ref{labyrinthSplit} for $split1a$ and $split2a$.

We need to apply the $split$ rules in this specific order or a possible split might be missed. For example, $split1a$ might be applicable somewhere where $split3a$ is also applicable but by applying $split1a$ first we would not then explore the labyrinth to the West or South. This could lead to ending up with a longer path to the exit or in the worst case not finding the exit at all. 

All $split$ rules have an empty $M$ subgraph. This will allow the strategy to move each $Pather$ at most once per iteration. The strategy will do this and then use Property($\rho$,G) to add all the $Pathers$ back to $P$ and then start over again. This ensures that no $Pather$ is given priority and is needed to find the shortest path (as explained further down).

\begin{figure}[h]
\centering
\includegraphics[width=0.4\columnwidth, keepaspectratio]{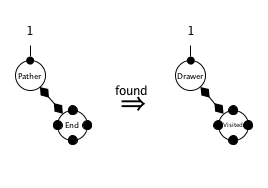}
\centering
\includegraphics[width=0.4\columnwidth, keepaspectratio]{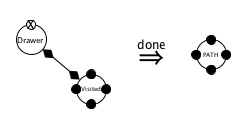}
\centering
\includegraphics[width=0.4\columnwidth, keepaspectratio]{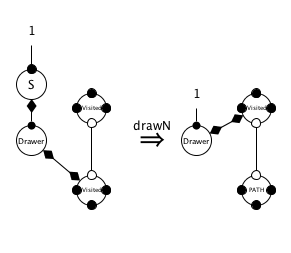}
\caption{The $found$, $done$ and  $drawN$ rules.}
\label{labyrinthFound}\label{labyrinthDone}\label{labyrinthDraw}
\end{figure}

While trying to apply the $split$ rules in that specific order, the strategy will constantly check if the $found$ rule (in Figure \ref{labyrinthFound} ) is applicable. If it is, it moves onto Step 2: drawing the path. If the $End$ node is not reachable from the starting point, the program will not terminate.

Step 2 checks if the $done$ rule (Figure \ref{labyrinthDone} ) is applicable and if it is not it will attempt to apply the four $draw$ rules $drawN$, $drawE$, $drawS$ and $drawW$. See Figure \ref{labyrinthDraw} for the $drawN$ rule; the other three $draw$ rules are similar and cater to a different direction each.

If the $done$ rule is applicable, the program will terminate and our labyrinth will have the shortest path to the exit drawn on it.

\begin{itemize}

\item LabStrat: $Step1$ ; $Step2$



\item  Step1: while(Not($found$))do(repeat$_*$($Step1Split$); Property(Y,G))min(0)max(-1); $found$
\item Y is: type=="Pather"

\item Step1Split: $split4$ orelse $split3a$ orelse $split3b$ orelse $split3c$ orelse $split3d$ orelse $split2a$ orelse $split2b$ orelse $split2c$ orelse $split2d$ orelse $split2e$ orelse $split2f$ orelse $split1a$ orelse $split1b$ orelse $split1c$ orelse $split1d$

\item Step2:  while(Not($done$))do($drawN$ orelse $drawE$ orelse $drawS$ orelse $drawW$)min(0)max(-1); $done$

\end{itemize}

Since \emph{Labyrinth}  agents are changed to $Visited$ agents when a $Pather$ moves from them, if a branching occurs in the labyrinth that later reconnects (as seen at the lower middle Figure \ref{labyrinthMap}) the branch with the shortest path will be picked (remembering that each $Pather$ can only take one step at most during each iteration so the $Pather$ in the shortest branch will get to the reconnecting \emph{Labyrinth} node first).

Branching that reconnects will cause stuck $Pathers$. When the $Pather$ from the quickest branch gets to the reconnecting \emph{Labyrinth} node, it will split and go to the slowest branch. That newly split $Pather$ will eventually meet the original $Pather$ of that branch (going the other way). These two $Pather$ agents will be positioned in two adjacent \emph{Labyrinth} agents but won't be able to move and remain stuck there.
We could extend our graph program by creating a set of rules to eliminate these stuck $Pather$ agents, using the $\epsilon$ agent. This is mainly an aesthetic improvement since the stuck $Pather$ agents will not affect the functionality of the program.

\section{Related Work: Graph Rewriting Tools}

\label{sec:related}

Several tools are available to edit graphs, and some of them allow
users to model graph transformations. Below we review some of the
systems that share some goals with PORGY.

GROOVE~\cite{Rensink03} is centered around the use of simple graphs
for modeling the design-time, compile-time, and run-time structure of
object-oriented systems. 
Visualisation is not the main objective,
and after each rewrite step the user must update the layout of the
graph by hand.
GROOVE permits to control the application of rules, via a control
language with sequence, loop, random choice, try()else() and simple
(non recursive) function calls.  These are similar to PORGY's
constructs, but GROOVE's language does not
include the notion of position; thus, it is not possible to specify 
directly a position for the application of rules.

The Fujaba~\cite{NickelNZ00} 
Tool Suite is an Open Source CASE tool providing developers with
support for model-based software engineering and re-engineering.
Fujaba has a basic strategy language, including conditionals, sequence
and method calls. There is no parallelism, and again one of the main
differences with PORGY is that Fujaba does not include a notion of
location to guide the rule application.

AGG~\cite{ErmelRT97} is a rule based visual language supporting an
algebraic approach to graph transformation. It aims at the
specification and prototypical implementation of applications with
complex graph-structured data. 
The application
of rules can be controlled by defining \emph{layers} and then
iterating through and across layers. Again, there is no notion of
position and there is no control on the search for redexes.

PROGRES~\cite{Schurr97b} offers  an executable
specification language based on graph rewriting systems (graph
grammars). The aim is to combine EER-like and UML-like class diagrams
for the definition of complex object structures with graph rewrite
rules for the definition of operations on these structures.  PROGRES
allows users to define the way rules are applied (it includes
non-deterministic constructs, sequence, conditional and looping) but
it does not allow users to specify the position where the rule is
applied. It is a very expressive language and includes a tracing
functionality through backtracking.

GrGen.NET~\cite{GeissBGHS06} is a programming tool for graph
transformation designed to ease the transformation of complex graph
structured data as required in model transformation, computer
linguistics, or modern compiler construction, for example. It is comparable to other
programming tools like parser generators which ease the task of formal
language recognition.


GP~\cite{Plump09} is a rule-based, non-deterministic programming
language. Programs are defined by sets of graph rewriting rules and a
textual expression that describes the way in which rules should be
applied to a given graph. 
The simplest expression is a set of rules, and this means that any of the rules can be applied to rewrite the graph. The language has three main control constructs: sequence, repetition and conditional (if-then-else), and it has been shown to be complete. 
It uses a built-in Prolog-like backtracking technique
(users cannot easily handle the derivation tree or change the backtracking algorithm).

%
%



GReAT (Graph Rewriting and Transformation)~\cite{BalasubramanianNBK06}
is a tool for building model transformation tools. 
Rule execution is sequential and there are conditional and
looping structures. 






PORGY and its strategy language allow a higher expressive power
with its focus on position. Strategies are not limited to picking
random applications but can travel through the graph in a dynamic and
strategic manner to apply rules and sub-strategies.  PORGY has also a
strong focus on visualisation and scale, thanks to the TULIP back-end
which can handle large graphs with millions of elements and comes with
powerful visualization and interaction features. Some of Tulip's built-in
functionalities, such as selecting a node in the trace for highlighting
its "lifetime" within the trace, give the user an immediate visual
feedback.

\section{Conclusion and Future Work}
\label{sec:conc}

The strategy language defined in this paper is part of the
PORGY~\cite{AndreiO:icgt} system, which is an environment that allows
users to define graphs and graph transformation rules. PORGY is
implemented using the TULIP platform, for the visualisation of graphs
and graph transformation rules.  PORGY provides also tools to
visualise traces of rewriting, and the strategy language is used in
particular to guide the construction of the traces.

Although PORGY and its strategy language were implemented specifically
to work with port graphs (and interaction nets in particular), the
strategy language could be applied to other graph formalisms (e.g.,
term graphs). This is a direction for future work. Verification and
debugging tools for avoiding conflicting rules or non-termination
 are also planned for future work.
The PORGY strategy language is not minimal and finding a set of minimal 
constructs will also be a subject for future work.

\paragraph{Acknowledgements} We are grateful to the members of the PORGY team, and in particular to H\'{e}l\`{e}ne Kirchner, for many inspiring discussions on the topics of this paper.

\label{sect:bib}

\bibliographystyle{eptcs} 
\bibliography{bib}

\end{document}